\documentclass[longauth]{aa}  
\usepackage{graphicx}
\usepackage{txfonts}
\usepackage{natbib}
\usepackage{subfigure}
\bibpunct{(}{)}{;}{a}{}{,}

\begin{document}

   \title{Radial decoupling of small and large dust grains in the transitional disk RX J1615.3-3255}

   \author{Robin Kooistra\inst{1}
           \and Inga Kamp\inst{1}
           \and Misato Fukagawa\inst{2}
           \and Francois M\'enard\inst{3}
           \and Munetake Momose\inst{4}
           \and Takashi Tsukagoshi\inst{4}
	   \and Tomoyuki Kudo\inst{5}
	   \and Nobuhiko Kusakabe\inst{6}
	   \and Jun Hashimoto\inst{6}
	   \and Lyu Abe\inst{7}
	   \and Wolfgang Brandner\inst{8}
	   \and Timothy D. Brandt\inst{9}
	   \and Joseph C. Carson\inst{10}
	   \and Sebastian E. Egner\inst{5}
	   \and Markus Feldt\inst{8}
	   \and Miwa Goto\inst{11}
	   \and Carol A. Grady\inst{12,13,14}
	   \and Olivier Guyon\inst{5}
           \and Yutaka Hayano\inst{5}
           \and Masahiko Hayashi\inst{15}
           \and Saeko S. Hayashi\inst{5}
           \and Thomas Henning\inst{8}
           \and Klaus W. Hodapp\inst{16}
           \and Miki Ishii\inst{15}
           \and Masanori Iye\inst{15}
           \and Markus Janson\inst{17}
           \and Ryo Kandori\inst{15}
           \and Gillian R. Knapp\inst{18}
           \and Masayuki Kuzuhara\inst{19}
           \and Jungmi Kwon\inst{20}
           \and Taro Matsuo\inst{21}
           \and Michael W. McElwain\inst{12}
           \and Shoken Miyama\inst{22}
           \and Jun-Ichi Morino\inst{15}
           \and Amaya Moro-Martin\inst{18,23}           
           \and Tetsuo Nishimura\inst{5}
           \and Tae-Soo Pyo\inst{5}
           \and Eugene Serabyn\inst{24}      
           \and Takuya Suenaga\inst{15,25}
           \and Hiroshi Suto\inst{15,6}
           \and Ryuji Suzuki\inst{15}
           \and Yasuhiro H. Takahashi\inst{20,15}
           \and Michihiro Takami\inst{26}
           \and Naruhisa Takato\inst{5}
           \and Hiroshi Terada\inst{15}
           \and Christian Thalmann\inst{27}
           \and Daigo Tomono\inst{5}
           \and Edwin L. Turner\inst{18,28}
           \and Makoto Watanabe\inst{29}
           \and John Wisniewski\inst{30}
           \and Toru Yamada\inst{31}
           \and Hideki Takami\inst{15}
           \and Tomonori Usuda\inst{15}
           \and Motohide Tamura\inst{20,15,6}
           \and Thayne Currie\inst{5}
           \and Eiji Akiyama\inst{15}
           \and Satoshi Mayama\inst{32,33}
           \and Katherine B. Follette\inst{28}
           \and Takao Nakagawa\inst{34}
           }

   \institute{Kapteyn Astronomical Institute, University of Groningen, Postbus 800, 9700 AV Groningen, The Netherlands
   \and Division of Particle and Astrophysical Science, Graduate School of Science, Nagoya University, Furo-cho, Chikusa-ku, Nagoya, Aichi 464-8602, Japan
   \and Univ. Grenoble Alpes, CNRS, IPAG, F-38000 Grenoble, France
   \and College of Science, Ibaraki University, 2-1-1 Bunkyo, Mito, Ibaraki 310-8512, Japan
   \and Subaru Telescope, 650 North A'ohoku Place, Hilo, HI 96720, USA
   \and Astrobiology Center of NINS, 2-21-1, Osawa, Mitaka, Tokyo, 181-8588, Japan
   \and Laboratoire Lagrange (UMR 7293), Universit\'e de Nice-Sophia Antipolis, CNRS, Observatoire de la C\^ote d'Azur, 28 avenue Valrose, 06108 Nice Cedex 2, France
   \and Max Planck Institute for Astronomy, K{\"o}nigstuhl 17, D-69117 Heidelberg, Germany
   \and Astrophysics Department, Institute for Advanced Study, Princeton, NJ, USA
   \and Department of Physics and Astronomy, College of Charleston, 58 Coming St., Charleston, SC 29424, USA
   \and Universit{\"a}ts-Sternwarte M{\"u}nchen, Ludwig-Maximilians-Universit{\"a}t, Scheinerstr. 1, 81679 Mu\''nchen, Germany
   \and Exoplanets and Stellar Astrophysics Laboratory, Code 667, Goddard Space Flight Center, Greenbelt, MD 20771, USA
   \and Eureka Scientific, 2452 Delmer, Suite 100, Oakland CA96002, USA
   \and Goddard Center for Astrobiology
   \and National Astronomical Observatory of Japan, Mitaka, Tokyo 181-8588, Japan
   \and Institute for Astronomy, University of Hawaii, 640 N. A'ohoku Place, Hilo, HI 96720, USA
   \and Department of Astronomy, Stockholm University, AlbaNova University Center, SE-106 91 Stockholm, Sweden
   \and Department of Astrophysical Science, Princeton University, Peyton Hall, Ivy Lane, Princeton, NJ 08544, USA
   \and Department of Earth and Planetary Sciences, Tokyo Institute of Technology, 2-12-1 Ookayama, Meguro-ku, Tokyo 152-8551, Japan
   \and Department of Astronomy, The University of Tokyo, 7-3-1, Hongo, Bunkyo-ku, Tokyo, 113-0033, Japan
   \and Department of Earth and Space Science, Graduate School of Science, Osaka University, 1-1 Machikaneyamacho, Toyonaka, Osaka 560-0043, Japan
   \and Hiroshima University, 1-3-2, Kagamiyama, Higashihiroshima, Hiroshima 739-8511, Japan
   \and Department of Astrophysics, CAB-CSIC/INTA, 28850 Torrej{\'o}n de Ardoz, Madrid, Spain
   \and Jet Propulsion Laboratory, California Institute of Technology, Pasadena, CA, 91109, USA
   \and Department of Astronomical Science, The Graduate University for Advanced Studies, 2-21-1, Osawa, Mitaka, Tokyo, 181-8588, Japan
   \and Institute of Astronomy and Astrophysics, Academia Sinica, P.O. Box 23-141, Taipei 10617, Taiwan
   \and Swiss Federal Institute of Technology (ETH Zurich), Institute for Astronomy,
   Wolfgang-Pauli-Strasse 27, CH-8093 Zurich, Switzerland
   \and Kavli Institute for Physics and Mathematics of the Universe, The University of Tokyo, 5-1-5, Kashiwanoha, Kashiwa, Chiba 277-8568, Japan
   \and Department of Cosmosciences, Hokkaido University, Kita-ku, Sapporo, Hokkaido 060-0810, Japan
   \and H. L. Dodge Department of Physics \& Astronomy, University of Oklahoma, 440 W Brooks St Norman, OK 73019, USA
   \and Astronomical Institute, Tohoku University, Aoba-ku, Sendai, Miyagi 980-8578, Japan
   \and The Center for the Promotion of Integrated Sciences, The Graduate University for Advanced Studies~(SOKENDAI), Shonan International Village, Hayama-cho, Miura-gun, Kanagawa 240-0193, Japan
   \and Department of Astronomical Science, The Graduate University for Advanced Studies ~ (SOKENDAI), 2-21-1 Osawa, Mitaka, Tokyo 181-8588, Japan
   \and Institute of Space and Astronautical Science, Japan Aerospace Exploration Agency, 3-1-1 Yoshinodai, Chuo-ku, Sagamihara, Kanagawa 252-5210, Japan
   }

   \date{\today}

  \abstract
   {We present H-band (1.6 $\mathrm{\mu m}$) scattered light observations of the transitional disk RX J1615.3-3255, located in the $\sim$1 Myr old Lupus association. From a polarized intensity image, taken with the HiCIAO instrument of the Subaru Telescope, we deduce the position angle and the inclination angle of the disk. The disk is found to extend out to 68 $\pm$ 12 AU in scattered light and no clear structure is observed. Our inner working angle of 24 AU does not allow us to detect a central decrease in intensity similar to that seen at 30 AU in the 880 $\mathrm{\mu m}$ continuum observations. We compare the observations with multiple disk models based on the Spectral Energy Distribution (SED) and submm interferometry and find that an inner rim of the outer disk at 30 AU containing small silicate grains produces a polarized intensity signal which is an order of magnitude larger than observed. We show that a model in which the small dust grains extend smoothly into the cavity found for large grains is closer to the actual H-band observations. A comparison of models with different dust size distributions suggests that the dust in the disk might have undergone significant processing compared to the interstellar medium.}
   

   \maketitle
%

\section{Introduction}\label{secintro}
The definition of transitional disks is heavily debated in the literature. Following the review by \citet{art:espaillat}, they are objects that exhibit almost no near-IR excess, yet harbor a strong mid- and far-IR excess. The former suggests that the inner regions have been cleared of material, forming a hole in the disk. \citet{art:strom} suggested that they are a transition stage in the evolution from an optically thick disk extending towards the star into a dispersed low-mass disk. Disks that do have a near-IR excess, but a dip in the mid-IR emission can be interpreted as a two-disk system with a gap in between \citep{art:espaillat2}. They are sometimes referred to as pre-transitional disks \citep{espaillat07}.\\
A key question for studying these objects is how their inner region is cleared out. Several mechanisms have been proposed for this. Due to viscosity in disks, they are expected to become optically thin as they accrete \citep{pp6}, but this is a slow process. Photoevaporation, where material on the surface of the disk is heated strongly by the UV or X-ray radiation from the central star, can cause an outflow of material from the disk \citep{pp6}. This cuts off the supply of disk material from the outer disk, making it possible to clear out the inner disk as the material quickly accretes onto the star. A substantial sample of transitional disks, however, show too large inner holes together with too high accretion rates to be explained by photoevaporation alone \citep{art:owen}. Furthermore, disk winds driven by magnetohydrodynamic turbulance can also play an important role in the dispersal of disks \citep{diskwinds}. It is also possible to create gaps in the disk by dynamical interaction with single \citep[e.g.][]{pinilla12,dejuanovelar13} or multiple massive objects \citep[e.g.][]{zhu11,dodsonrobinson11,dong15}. The main candidate for this type of clearing is a planetary body, carving a hole by sweeping up material as it moves through the disk. The transitional disks are thought to be an important stage in understanding the formation of these planets and several promising canditates for planets within disks have been observed, e.g. HD 100546 \citep{hd100546,hd100546_2}, HD 169142 \citep{hd169142}, LkCa 15 \citep{lkca15_2,lkca15}.\\
A significant sample of transitional disks are shown to have a cavity in the submm \citep{art:andrews}. Recently however, comparisons of high resolution near-IR observations with submm images have shown a possible number of transitional disks with near-IR emission extending into the submm cavity \citep[e.g.][]{dong12}, which suggests a decoupling of the distribution of both small and large dust grains. In some cases the cavities at both wavelengths have been spatially resolved, indeed showing that the small dust grains can move in closer to the star than the large dust grains \citep[e.g.][]{muto12,garufi,grady13,follette13}.\\
The transitional disk RX J1615.3-3255 (from here on referred to as RXJ1615) was first detected by \citet{art:henize}. RXJ1615 is located in the constellation Lupus and has been kinematically tied to a young ($\sim$1 Myr old) subgroup of the Lupus association at a distance of 185 pc \citep{art:makarov}. It was identified as a weak-line T Tauri star by \citet{art:krautter}, based on optical spectroscopy. Spitzer IR observations, showed the first evidence of an inner hole in the disk by performing a fit to the SED, designating it a transitional disk \citep{merin}.\\
\citet{art:andrews} performed high resolution 880 $\mathrm{\mu m}$ observations and created a disk model that fits both the spectral energy distribution (SED) and their visibilities. Their data shows a decrease in the intensity close to the center, a low density cavity out to a radius of 30 AU. The disk is relatively flat and massive, $\sim 12 \%$ of the stellar mass. Because of the large size of the gap and the disk mass, they find that the disk is most likely cleared dynamically by tidal interactions with a low-mass brown dwarf or giant planet companion on a long-period orbit.\\
In this paper, we present new H-band scattered light images of RXJ1615 and reproduce the disk model from \citet{art:andrews} based on submm interferometry and the SED in order to study the spatial distribution of the large versus small dust grains in this disk.

\section{Observations and Data Reduction}\label{secobs}
Our observations were obtained with the HiCIAO instrument \citep{tamura} of the Subaru telescope as part of the 16th run of the Strategic Explorations of Exoplanets and Disks with Subaru (SEEDS) survey on July 5th 2012. The images were taken in the H-band (1.6 $\mathrm{\mu m}$) using a combination of HiCIAO's quad Polarimetric Differential Imaging (qPDI) and Angular Differential Imaging (ADI) modes. In order to explore structures close to the star, no coronographic mask was used. A summary of the observations and the details of the data are given in Table \ref{tab:obs}.\\

\begin{table}[ht!]
 \caption{Summary of the observations and the obtained data}
 \label{tab:obs}
 \centering
 \begin{tabular}{l l l l}
 \hline
 Object details & & &\\
 \hline
 Name & \multicolumn{3}{l}{RX J1615.3-3255}\\
 $\alpha$[J2000] & \multicolumn{3}{l}{16 15 20.20}\\
 $\delta$[J2000] & \multicolumn{3}{l}{-32 55 05.1}\\
 Region & \multicolumn{3}{l}{Lupus ($\sim$ 1 Myr)}\\
 Distance & \multicolumn{3}{l}{185 pc}\\
 Spectral type & \multicolumn{3}{l}{K5}\\
 \hline
 Observation conditions & & &\\
 \hline
 Telescope, instrument & \multicolumn{3}{l}{Subaru, HiCIAO}\\
 Observation modes & \multicolumn{3}{l}{qPDI $\&$ ADI}\\
 Observed band & \multicolumn{3}{l}{H-band (1.6 $\mathrm{\mu m}$)}\\
 Diffraction limit & \multicolumn{3}{l}{0.05 arcsec}\\
 Pixel scale & \multicolumn{3}{l}{$9.5\cdot10^{-3}$ arcsec}\\
 Standard star & \multicolumn{3}{l}{HD203856}\\
 R-band mag observed & \multicolumn{3}{l}{11.6}\\
 Seeing & \multicolumn{3}{l}{> 1"}\\
 \hline
 Data & & & \\
 \hline
 Final integration time & \multicolumn{3}{l}{7.5 minutes}\\
 & Date & $N_{\mathrm{exp}}$ & $t_{\mathrm{exp}}$(s)\\
 \cline{2-4}
 Star & 2012-07-05 & 68 & 30\\
 Standard star & & 3 & 1.5\\
 Flat field& 2012-05-13 & 9 & 120\\
 Dark & 2012-07-07 & 100 & 40\\
 \hline
 \end{tabular}
 \tablefoot{$N_{\mathrm{exp}}$ is the number of exposures and $t_{\mathrm{exp}}$ is the exposure time for each exposure.}
\end{table}

\subsection{Data reduction}\label{secdata}
We performed the data reduction following the standard description for handling PDI data to obtain the Stokes parameters using cycles of four different angular positions of the Wollaston prism \citep{art:speck}. All images were destriped, corrected for warm and bad pixels, distortion and instrumental polarization. The images were then derotated to account for the ADI observation mode and photometrically callibrated using data of the standard star HD203856 observed in the open use program on July 5th 2012. 

\subsection{Bad images}\label{secbadim}
\begin{figure}
 \resizebox{\hsize}{!}{\includegraphics{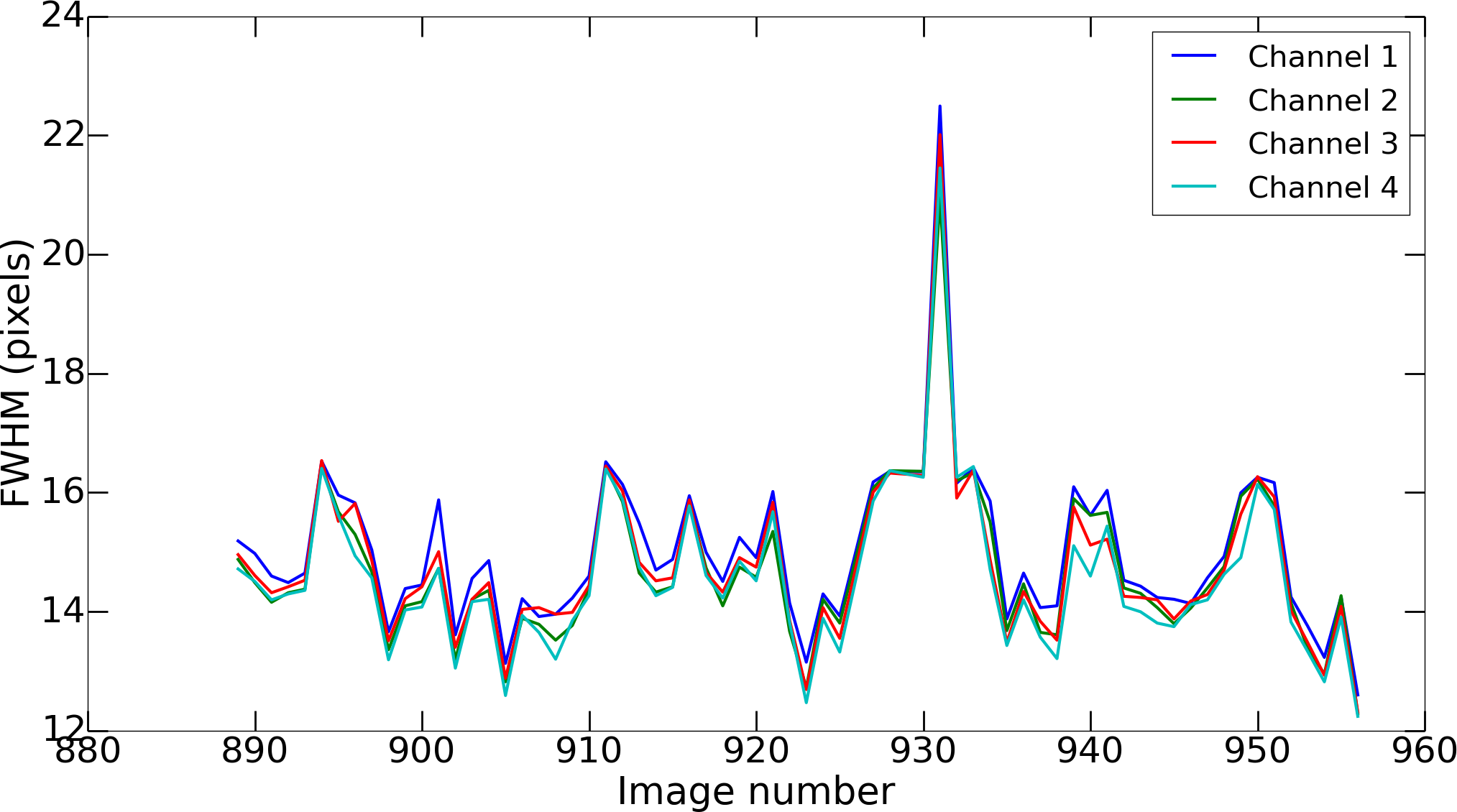}}
 \caption{The FWHM in different images. Each line represents a different qPDI channel.}
 \label{fig:fwhm}
\end{figure}

During the data reduction process, we noticed that the quality between exposures varied. This is evidenced in Fig. \ref{fig:fwhm}, where we show the FWHM of the stellar light after the distortion correction step. Although the atmospheric attenuation during the night should not have been a problem, the seeing did get above 1" for a large part of the night. This is difficult for the adaptive optics to correct for, possibly explaining the variation in the point spread function (PSF) of the exposures. In order to try to resolve the cavity found by \citet{art:andrews}, we wanted to keep our inner working angle as small as possible. We therefore removed all data where the PSF FWHM is larger than 16 pixels (0.152 arcsec). This effectively meant losing half of the data, bringing the integration time down from 7.5 (15 waveplate cycles) to 4 minutes (8 waveplate cycles), significantly reducing our signal to noise, but decreasing our inner working angle from $\sim$33 AU to $\sim$24 AU.

\section{Results}\label{secres}
\begin{figure*}[ht!]
 \resizebox{0.5\hsize}{!}{\subfigure{\includegraphics{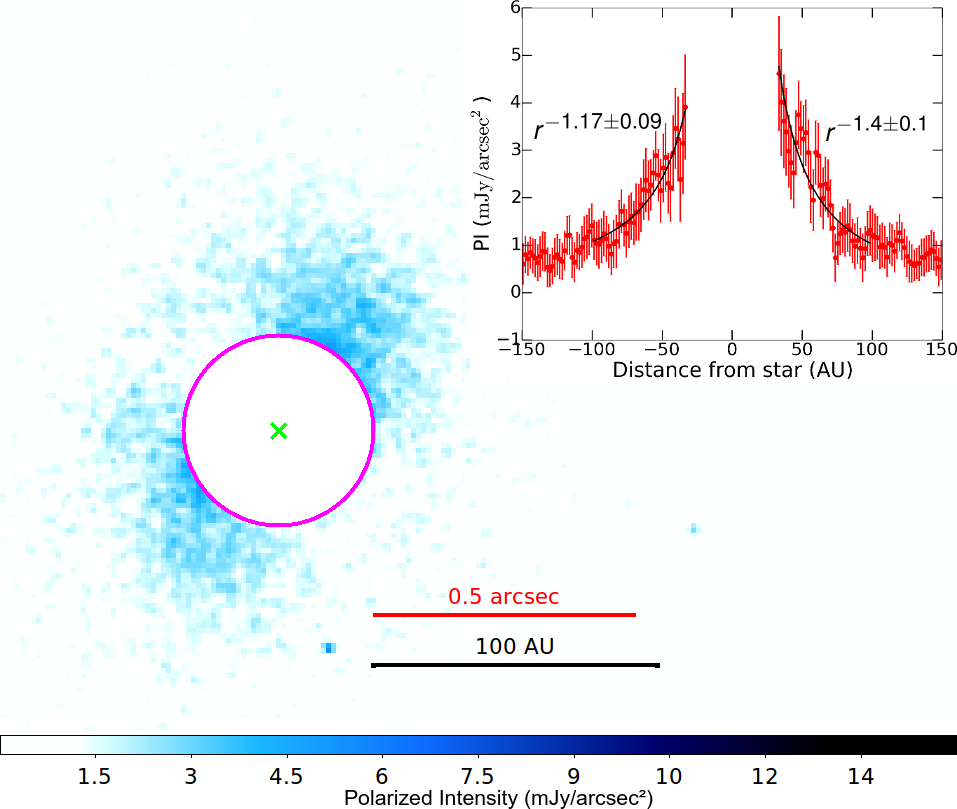}}}
 \resizebox{0.5\hsize}{!}{\subfigure{\includegraphics{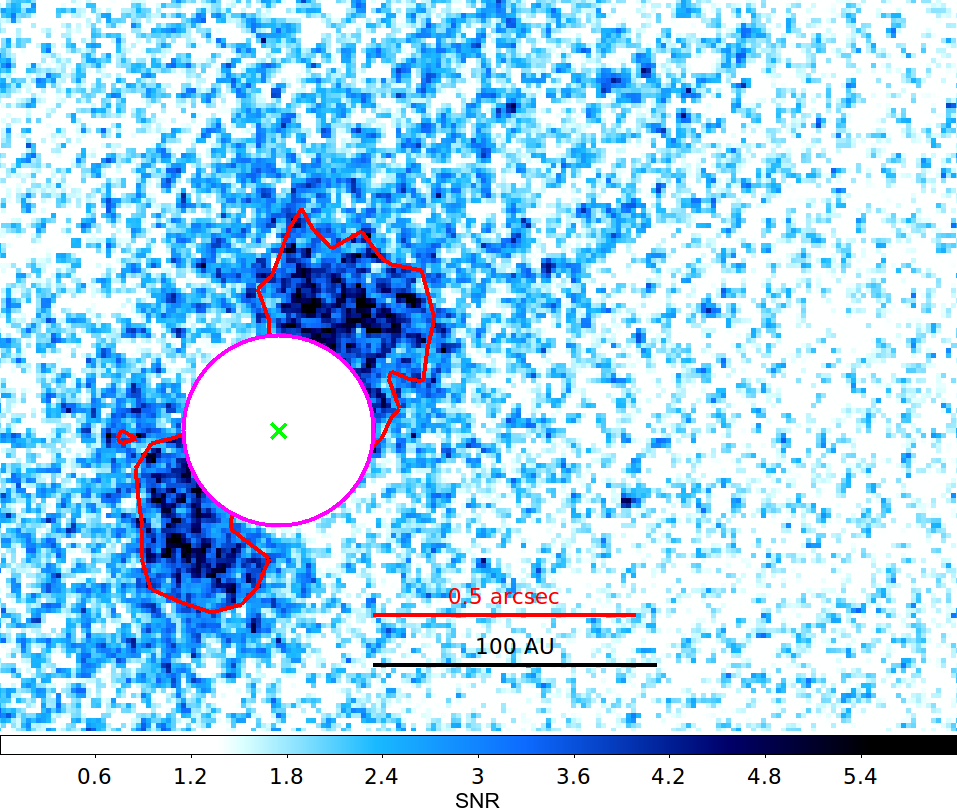}}}
 \caption{Polarized intensity map (left) and signal-to-noise map (right). The red contours show the SNR > 3 region, the green cross denotes the stellar position and the magenta circle gives the inner working angle of 33 AU. We mask the region inside the inner working angle. The plot gives the brightness profile along the disk major axis to which we fitted power-law profiles (black lines). The fitted power-law exponents are given in the plot with 1-$\sigma$ uncertainties.}
 \label{fig:pi}
\end{figure*}

Using the Stokes Q and U images obtained from the data, we determined the polarized intensity using
\begin{equation}
 PI = \sqrt{Q^2 + U^2}\label{eq:pi}
\end{equation}
The resulting polarized intensity map is given in the left panel of Fig \ref{fig:pi}, where the brightness profile along the major axis of the disk is given inside the plot. Power-law fits to both sides are shown in the plot as black lines. The disk has a relatively shallow radial profile along the major axis, with power-law indices of $-1.17\pm0.09$ and $-1.4\pm0.1$, compared to the range of -1.7 to -5 found in other disks \citep{kusakabe12,muto12}. As a conservative estimate for the errors on the polarized intensity we used the standard deviation between images of different waveplate cycles, divided by the square root of the number of images. Dividing the polarized intensity by the error then gives the signal-to-noise map shown in the right panel of Fig. \ref{fig:pi}. As can be seen in both figures, we clearly detected the extended disk emission with a signal-to-noise of $\sim$2-6.  

From the polarized intensity image we estimated a few physical parameters of the RXJ1615 disk and summarize them in Table \ref{tab:meas}. For the position angle (PA), we fitted an ellipsoid using the Image Reduction and Analysis Facility (IRAF\footnote{IRAF is distributed by National Optical Astronomy Observatory, which is operated by the Association of Universities for Research in Astronomy, Inc., under cooperative agreement with the National Science Foundation.}) package. Assuming an infinitely flat disk, we got the inclination angle ($i$) from the ratio of the major and minor axis of the disk. The uncertainties in these parameters are determined by the IRAF ellipsoid fitting routine. We defined the outer radius ($R_{\mathrm{out}}$) as the distance from the stellar position along the major axis at which the SNR drops below 3. Note that the outer radius of 68 $\pm$ 12 AU is smaller than what was found from the submm observations of \citet{art:andrews} ($\sim$115 AU). At large distances, the low surface brightness of the disk can be expected to disappear into the noise and therefore our measurement is likely underestimating the true outer radius of the disk. Our inner working angle of 33 AU does not allow for the exploration of the inner 30 AU cavity. However, when decreasing the inner working angle to 24 AU by removing the bad images, as described in section \ref{secbadim}, we still do not see signs of a depletion in the inner 30 AU of the disk. This could signify that, although the large dust grains ($\sim$mm size) are depleted in the gap (as evidenced by the submm data), the small dust grains ($\lesssim$ 1 $\mu$m size) either still survive, or they have a smaller cavity size than the large grains. The latter was also seen in other transitional disks, e.g. SAO 206462 \citep{muto12,garufi}, MWC 758 \citep{grady13}, SR21 \citep{follette13}. However, our inner working angle of 24 AU is too close to the 30 AU cavity radius of \citet{art:andrews} to be certain from this dataset.

\begin{table}[ht!]
 \caption{Physical parameters measured using an ellipsoid fit.}
 \label{tab:meas}
 \centering
 \begin{tabular}{l l}
 \hline
 Parameter (units) & Value\\
 \hline
 PA (degrees)\tablefootmark{a} & $142\pm1$\\
 $i$ (degrees)\tablefootmark{b} & $52.91\pm0.02$\\
 $R_{\mathrm{out}}$ (AU) & $68\pm12$\\
 \hline
 \end{tabular}
 \tablefoot{We give 1-$\sigma$ errors.
 \tablefoottext{a}{Major axis, measured East from the North.}
 \tablefoottext{b}{$0^\circ$ is face-on.}}
\end{table}

In order to confirm the nature of the scattered light, one can look at the direction of the polarization vector. The polarization angle is defined as follows
\begin{equation}
 \alpha = \frac{1}{2}\tan^{-1}\left(\frac{U}{Q}\right)\label{eq:polang}
\end{equation}

\begin{figure}[ht!]
 \resizebox{\hsize}{!}{\includegraphics{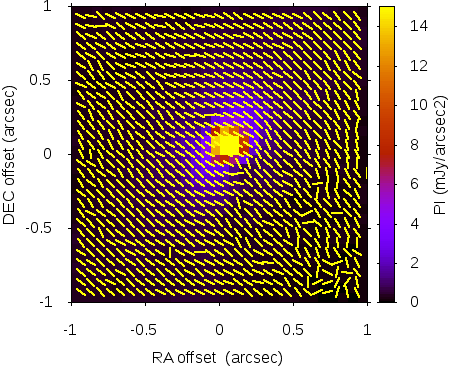}}
 \caption{Polarization angle map overlayed on a block averaged version of PI. The angle of the yellow ticks gives the polarization angle.}
 \label{fig:polang}
\end{figure}

We overplot this angle on the polarized intensity in Fig. \ref{fig:polang}, where the angle of the ticks denotes the polarization angle. Because in the disk region along the major axis most ticks are aligned in a direction perpendicular to the direction towards the stellar position, this is a clear sign that we are indeed looking at light scattered from the star through the disk. Note however, that the polarization angle is not aligned perpendicular to the radial direction along the minor axis of the disk and we thus might be suffering from the effects of a polarized halo \citep{hashimoto12}. This could affect our derivation of the outer radius, position angle and inclination of the disk and could also explain why we find a significantly larger inclination angle than \citet{art:andrews}.\\

\section{Comparison with Models}\label{secmod}
\citet{art:andrews} created a disk model of RXJ1615 that fits both the spectral energy distribution (SED) and their observed submm visibilities. In order to see if this same model also agrees with the scattered light observations, we used the Monte-Carlo three-dimensional continuum radiative transfer code MCFOST \citep{art:mcfost} to reproduce the \citet{art:andrews} disk model and simulate an H-band image. The code traces the path of individual packages of photons that propagate through the disk. The photons can undergo scattering, absorption and re-emission events. The main sources of radiation are thermal emission from dust in the disk and photospheric emission from the star. The thermal emission is assumed to be isotropic and depends only on the temperature, density and opacity of the disk material. The stellar emission is governed by the stellar photospheric spectrum. Photon packages that manage to escape the computation grid are used to calculate the SED and create the simulated H-band image.

\subsection{Disk models}\label{secmod2}
\begin{figure}[ht!]
 \resizebox{\hsize}{!}{\includegraphics{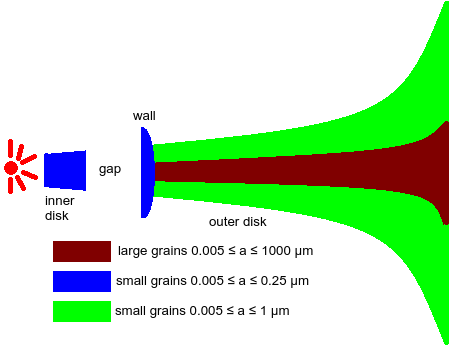}}
 \caption{A sketch of \citet{art:andrews} model.}
 \label{fig:sketch}
\end{figure}

The \citet{art:andrews} model has three main zones consisting of an inner and outer disk and a puffed up wall at the rim of the outer disk with a gap between the inner and outer disk that has been cleared of material (see Fig. \ref{fig:sketch}).\\
The density structure of the disk follows a Gaussian profile,
\begin{equation}
 \rho(r,z) = \rho_0(r)\,\mathrm{e}^{-\frac{z^2}{2H(r)^2}}
\end{equation}
where $H$ denotes the scale height. We assume that the dust follows the same vertical distribution as the gas. The disk flaring is characterized by the flaring exponent $\beta$
\begin{equation}
 H(r) = H_0\left(\frac{r}{r_0}\right)^\beta
\end{equation}
where $r_0$ is a reference radius and $H_0$ the corresponding scale height at that radius. The surface density profile follows either a power-law distribution (inner disk)
\begin{equation}
 \Sigma(r) = \Sigma_0\left(\frac{r}{r_0}\right)^{-\epsilon}
\end{equation}
or a tapered edge distribution (outer disk)
\begin{equation}
 \Sigma(r) = \Sigma_\mathrm{c}\left(\frac{r}{R_\mathrm{c}}\right)^{-\epsilon} \mathrm{e}^{-\left(\frac{r}{R_\mathrm{c}}\right)^{2-\epsilon}}\label{eq:tapered}
\end{equation}
characterized by the surface density exponent $\epsilon$. $R_\mathrm{c}$ in Eq. (\ref{eq:tapered}) denotes the characteristic radius at which the exponential term in the distribution becomes important. The dust grain population follows a power-law distribution in grain size ranging from the smallest size $a_{\mathrm{min}}$ to the largest $a_{\mathrm{max}}$ as $n(s)\propto a^{-3.5}$ and we use Draine astronomical silicates \citep{drainesil}.\\
In the outer disk, we allow for dust settling of the large dust grains onto the midplane of the disk \citep{dustsettl}. To mimick this effect, we assume two separate dust populations with different reference scale heights. The flaring index is kept constant throughout the disk. In order to reproduce the SED with MCFOST, we had to tweak some parameters compared to the ones adopted by \citet{art:andrews}, e.g. 150 K lower $T_{\mathrm{eff}}$, the inner disk dust mass was increased by a factor of 10 and the wall dust mass was decreased by a factor of 100. The full details of the model are given in Table \ref{tab:model}. To investigate the effect of different small dust grain distributions on the produced scattered light, we also explore two other models. One is the same as the \citet{art:andrews} model, but without the wall of intermediate sized dust grains at the inner rim of the outer disk. The other is a thin single disk model that fits the SED (no effort was made to fit details such as the Si feature at 10 micron). In this model the scale height is similar to that of the outer disk in the \citet{art:andrews} model, but the disk extends closer to the star and small and large dust grains are well mixed throughout the disk. The parameters of this model can also be found in Table \ref{tab:model}. All three resulting SEDs are shown in Fig. \ref{fig:sed}, where the photometric fluxes (see Table \ref{tab:phot} in the appendix) were de-reddened using the CCM law \citep{art:red} with $A_V$ = 0.4 and $R_V$ = 5. 

\begin{figure}[ht!]
 \resizebox{\hsize}{!}{\includegraphics{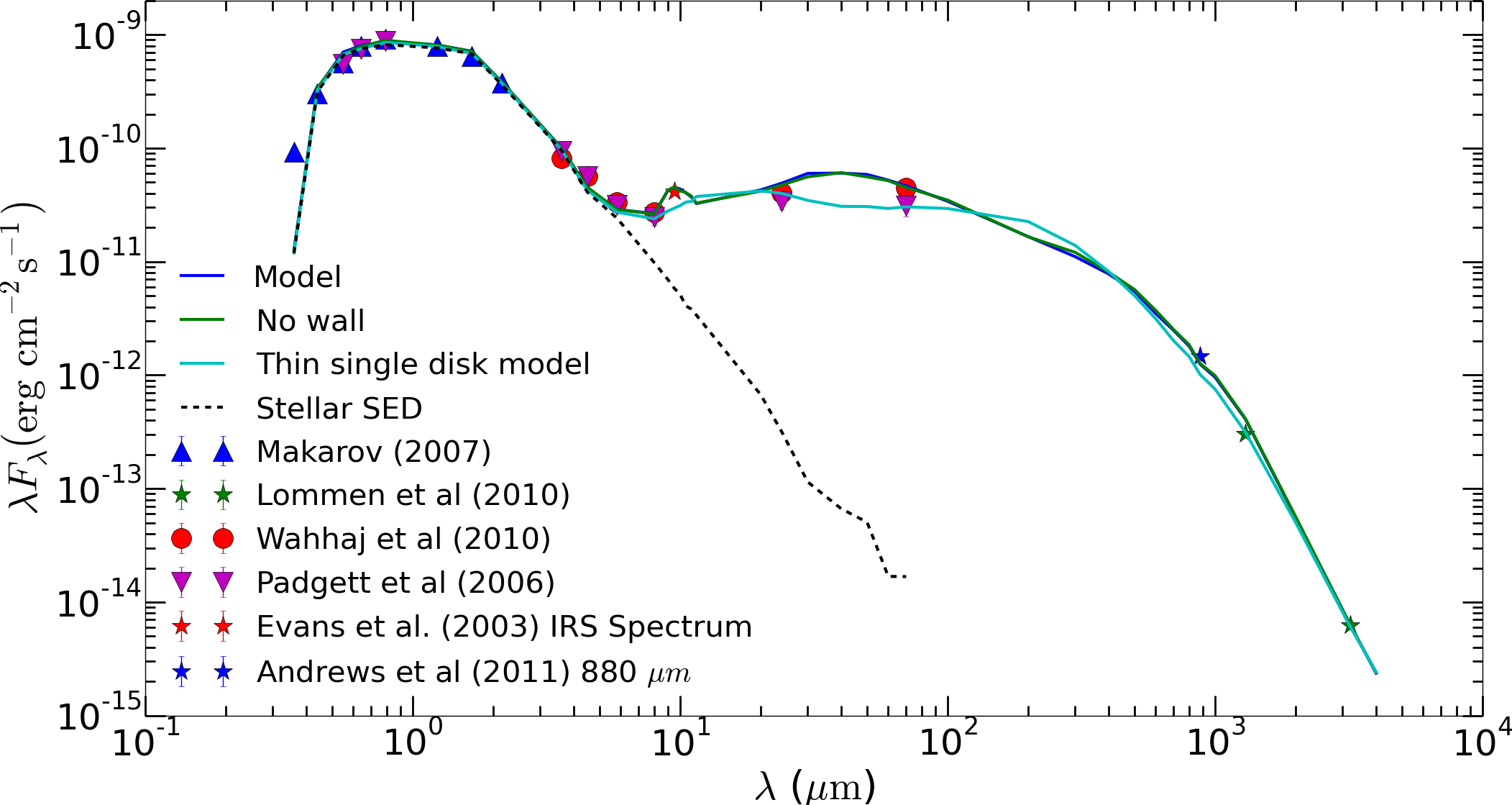}}
 \caption{SED of the disk models. The black dashed line shows the stellar SED from the output of MCFOST. The model parameters are given in Table \ref{tab:model}. The references for the photometric data points from the literature are given in the legend. The values of the uncorrected fluxes from the literature can be found in Table \ref{tab:phot} in the appendix.}
 \label{fig:sed}
\end{figure}

\begin{table*}[ht!]
 \caption{Model parameters}
 \label{tab:model}
 \centering
 \begin{tabular}{l l l l l l}
 \hline
 Stellar parameters & & & & &\\
 \hline
 $T_{\mathrm{eff}}$ & \multicolumn{5}{l}{4200 K}\\
 $R_{\ast}$ & \multicolumn{5}{l}{2.16$R_{\odot}$}\\
 $M_{\ast}$ & \multicolumn{5}{l}{1.10$M_{\odot}$}\\
 \hline
 Disk parameters & & & & &\\
 \hline
 Inclination angle & \multicolumn{5}{l}{$41.4^{\circ}$}\\
  & Inner disk & Wall & Outer disk, small grains & \multicolumn{1}{l|}{Outer disk, large grains} & Thin single disk model\\
 \cline{2-6}
 $\Sigma(r)$ distribution type & Power-law & Power-law & Tapered edge & \multicolumn{1}{l|}{Tapered edge} & Power-law\\
 $M_{\mathrm{dust}}$($M_{\odot}$) & $2.06\cdot10^{-9}$ & $1.10\cdot10^{-8}$ & $1.90\cdot10^{-4}$ & \multicolumn{1}{l|}{$1.08\cdot10^{-3}$} & $7\cdot10^{-4}$\\
 $H_{\mathrm{100 AU}}$(AU) & 3.4 & $H_{\mathrm{30 AU}}$ = 2.0 & 3.4 & \multicolumn{1}{l|}{0.68} & 3.5\\
 $R_{\mathrm{in}}$(AU) & 0.5 & 30 & 30.1 & \multicolumn{1}{l|}{30.1} & 2.7\\
 $R_{\mathrm{out}}$ or $R_{\mathrm{c}}$(AU) & 10 & 30.1 & 115 & \multicolumn{1}{l|}{115} & 115\\
 $\beta$ & 1.25 & 1.25 & 1.25 & \multicolumn{1}{l|}{1.25} & 1.3\\
 $\epsilon$ & 1 & 1 & 1 & \multicolumn{1}{l|}{1} & 1.5\\
 $a_{\mathrm{min}}$($\mathrm{\mu m}$) & 0.005 & 0.005 & 0.005 & \multicolumn{1}{l|}{0.005} & 0.005\\
 $a_{\mathrm{max}}$($\mathrm{\mu m}$) & 0.25 & 0.25 & 1 & \multicolumn{1}{l|}{1000} & 1000\\
 \hline
\end{tabular}
\tablefoot{$T_{\mathrm{eff}}$ Effective temperature; $R_\ast$ Stellar radius; $M_{\ast}$ Stellar mass; $M_{\mathrm{dust}}$ dust mass; $H_{\mathrm{100AU}}$ Scale height at 100 AU; $R_{\mathrm{in}}$ Inner radius; $R_{\mathrm{out}}$ Outer radius; $R_\mathrm{c}$ characteristic radius tapered edge distribution; $\beta$ Flaring exponent; $\epsilon$ surface density exponent; $a_{\mathrm{min}}$ Minimum size of the dust grains; $a_{\mathrm{max}}$ Maximum size of the dust grains.}
\end{table*}

\subsection{Simulated H-band Image}\label{secsimim}
\begin{figure}[ht!]
 \resizebox{\hsize}{!}{\includegraphics{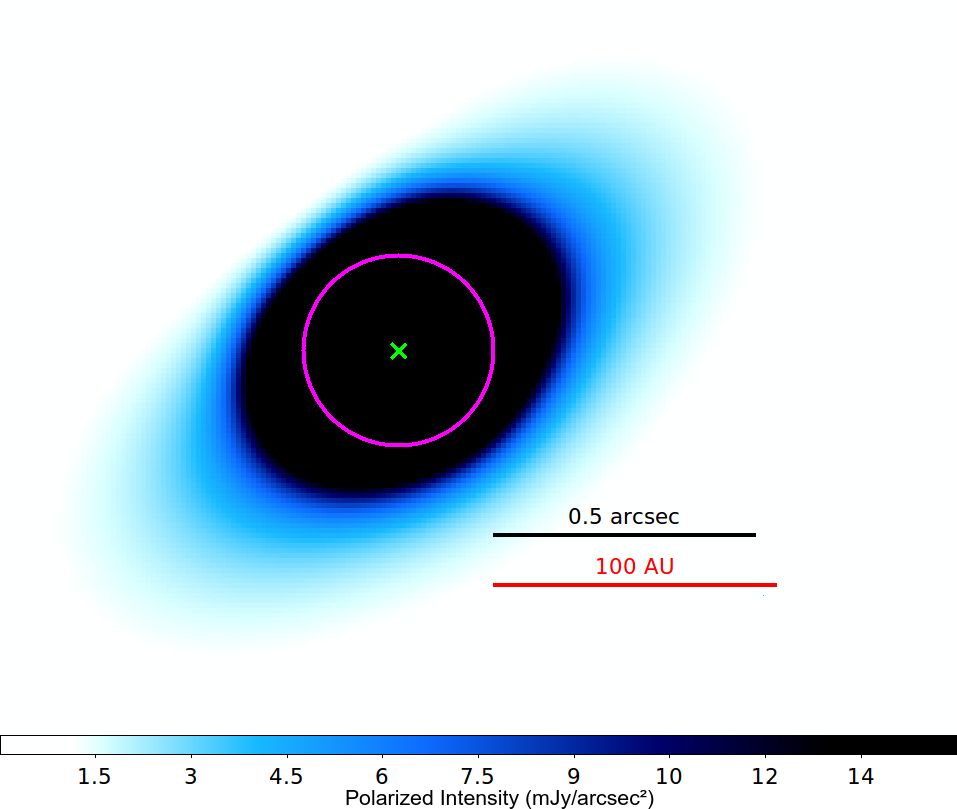}}
 \caption{Simulated H-band Polarized Intensity image of the \citet{art:andrews} model convolved with a two component Gaussian PSF with FWHM = 0.183 arcsec for the broad one and FWHM = 0.038 arcsec for the narrow one. The scales are identical to Fig. \ref{fig:pi} for comparison. The green cross gives the stellar position and the magenta circle denotes the inner working angle from our observations.}
 \label{fig:modpi}
\end{figure}

From MCFOST, we get Stokes Q and U images for the three disk models, which we converted to polarized intensity images of the model as it would be seen at a distance of 185 pc. Convolving this image with a point spread function (PSF) then allows for a better comparison of the model with the data. There is no data available of a PSF reference star observed on the same night, preventing an accurate PSF model. To still enable a qualitative comparison, we used the polarized intensity profile along the minor axis of the disk from our observations as a PSF estimate. This assumes that the disk is unresolved along the minor axis. We then fitted a double Gaussian to the mean of both sides of this profile (a broad component with FWHM = 0.184 arcsec and a narrow component with FWHM = 0.039 arcsec). The resulting convolved H-band image of the \citet{art:andrews} model is shown in Fig. \ref{fig:modpi}. 

\begin{figure}[ht!]
 \resizebox{\hsize}{!}{\includegraphics{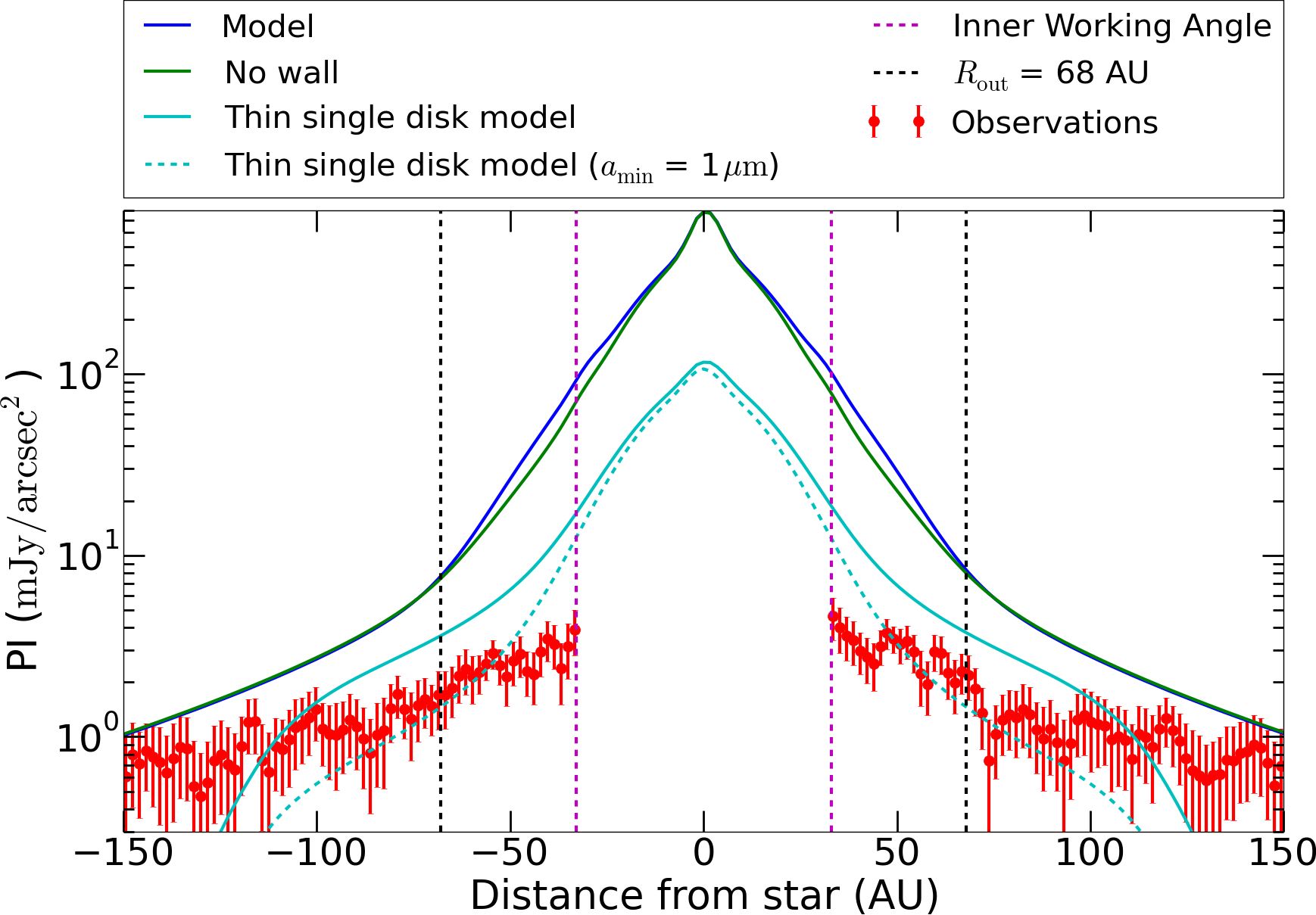}}
 \caption{Radial profile along the major axis of the simulated disk H-band PI images for the different models compared to our observations.}
 \label{fig:radprof}
\end{figure}

The extent of the emission is similar to our observations ($\sim$ 100 AU), but the model clearly produces a much higher signal. This is also evident from the radial profile of the polarized intensity along the major axis of the disk (Fig. \ref{fig:radprof}). Because the wall has a large surface area, it can be expected to produce a lot of scattered light \citep{thalmann10}. As can be seen from the radial profile, however, removing the wall only slightly lowers the polarized intensity signal. Instead, most of the emission comes from the inner rim at the edge of the outer disk (30 AU) and is then smeared out by the PSF. The thin single disk model reduces the polarized intensity by a factor of $\sim$ 2-4, bringing it down to a factor of a few from the observed polarized intensity in the outer $\sim$50 AU of the disk. Even though this model has almost the same scale height as the outer disk in the \citet{art:andrews} model, its inner radius is a lot smaller (2.7 AU vs 30 AU), also making the exposed surface area at the rim smaller ($H$ = 0.03 AU vs 0.76 AU) and thus reducing the scattered light intensity. This suggests that, although the large grain distribution in RXJ1615 has a cavity out to 30 AU consistent with the 880 $\mathrm{\mu m}$ observations, the small dust grains likely extend closer to the star.\\
However, this does not have to be the only option for bringing down the polarized intensity. A large inner rim much closer to the star could potentially cast a shadow and thus prevent the stellar light from scattering on the disk behind it \citep{dong12}. Such a wall would have to be massive enough, because, as we can see from our models, most of the light passes directly through the wall and scatters on the rim of the disk behind it instead. A more massive wall would block more stellar light, but would also affect the shape of the SED. Another possibility would be to change the properties of the dust grains (e.g. lowering the grain albedo or different dust type mixtures) or reducing the amount of small particles in the disk. The thin single disk model simply uses astronomical silicate grains with a minimum grain size of 0.05 $\mu$m, which give an albedo of 0.75 and a polarisability of 0.15. Changing $a_{\mathrm{min}}$ affects both these quantities. As can be seen from the dashed line in Fig \ref{fig:radprof}, changing the minimum dust grain size to $a_{\mathrm{min}}$ = 1 $\mu$m lowers the polarized intensity of the thin single disk model to the same level as we observe. In this case both the albedo and the polarisability decrease to 0.63 and 0.12, respectively. However, when increasing the minimum grain size even further to 3 $\mu$m, the albedo keeps decreasing as expected (0.57), but the polarisability rises again to much higher values (0.66), therefore increasing the disk polarized intensity again above observed values. The current (low) level of polarised intensity suggests that the dust in the disk has already undergone significant processing compared to the ISM. However, more observations, at different frequencies, and modelling are required before firm conclusions can be drawn.

\section{Conclusion}\label{secconcl}
We presented the first H-band scattered light observations of the transitional disk RX J1615.3-3255. We detected the disk in scattered light, finding an outer radius of 68 $\pm$ 12 AU. Outside our 24 AU inner working angle, we find no signs of the central depletion in the disk that was previously found in the submm. This could suggest a smaller cavity size for the small grains, but we suffer from low signal-to-noise and a large inner working angle. A detailed comparison with multiple disk models based on fits to the SED and submm visibilities does suggest that the small dust grain population is radially decoupled from the large grains. The small dust grains appear to be present closer to the star than the large grains and the dust in the disk has possibly undergone significant processing compared to the ISM. Future higher spatial resolution and higher sensitivity observations (e.g. VLT SPHERE) are required to provide more detailed information on the distribution of the small dust grains in this disk.

\begin{acknowledgements}\label{ackn}
We thank the Netherlands Foundation for Scientific Research support through the VICI grant 639.043.006. FMe acknowledges funding from ANR of France under contract number ANR-16-CE31-0013.
\end{acknowledgements}

\bibliographystyle{aa}
\bibliography{RXJ1615_biblio.bib}

\begin{appendix}
\section{Photometric table}\label{append}
\begin{table*}[ht!]
 \caption{Photometric data from the literature uncorrected for reddening. For some data points two references are available with slightly different calibration methods. We include both values for completeness.}
 \label{tab:phot}
 \centering
 \begin{tabular}{l l l l l l}
 \hline
 Wavelength ($\mathrm{\mu m}$) & Flux (mJy) & 1-$\sigma$ error (mJy) & Magnitude & Zero-magnitude Flux (Jy) & References\\
 \hline
 0.36 (U-band) & 5.1 & 0.2 & 13.88 & 1810 & 1,7\\
 0.44 (B-band) & 22.0 & 0.4 & 13.22 & 4260 & 1,7\\
 0.55 (V-band) & 57.7 & 0.5 & 12.00 & 3640 & 1,7\\
 0.55 (V-band) & 55.6 & 1.5 & 12.04 & 3640 & 2,7\\
 0.64 ($R_c$-band) & 101.1 & 0.9 & 11.21 & 3080 & 1,7\\
 0.64 ($R_c$-band) & 94.7 & 2.6 & 11.28 & 3080 & 2,7\\
 0.79 ($I_c$-band) & 160.9 & 1.5 & 10.50 & 2550 & 1,7\\
 0.79 ($I_c$-band) & 155.1 & 4.3 & 10.54 & 2550 & 2,7\\
 1.235 (J-band) & 268.2 & 5.9 & 9.435 & 1594 & 1,8\\
 1.662 (H-band) & 315.9 & 6.7 & 8.777 & 1024 & 1,8\\
 2.159 ($K_s$-band) & 251.6 & 4.4 & 8.558 & 666.7 & 1,8\\ 
 3.6 & 98.0 & 4.9 &  &  & 3\\
 3.6 & 114.0 & 17.1 &  &  & 2\\
 4.5 & 85.0 & 4.3 &  &  & 3\\
 4.5 & 85.0 & 12.8 &  &  & 2\\
 5.8 & 65.0 & 3.3 &  &  & 3\\
 5.8 & 61.0 &  9.2 &  &  & 2\\
 8.0 & 73.0 & 3.7 &  &  & 3\\
 8.0 & 66.0 & 9.9 &  &  & 2\\
 9.5 (Spitzer IRS spectrum) & 133.9 & 2.3 &  &  & 4\\
 24 & 322.0 & 32.2 &  &  & 3\\
 24 & 271.0 & 40.7 &  &  & 2\\
 70 & 1049.0 & 167.0 &  &  & 3\\
 70 & 727.0 & 145.4 &  &  & 2\\
 880 & 430.0 & 2.8 &  &  & 5\\
 1300 & 132.0 & 3.9 &  &  & 6\\
 3200 & 6.7 & 0.6 &  &  & 6\\
 \hline
\end{tabular}
\tablebib{
(1)~\citet{art:makarov}; (2) \citet{art:padgett}; (3) \citet{art:wahhaj}; (4) \citet{art:evans}; (5) \citet{art:andrews}; (6) \citet{art:lommen}; (7) \citet{art:bessel}; (8) \citet{art:2MASS}.
}
\end{table*}
\end{appendix}
\end{document}